# Experimental Demonstration of Optically Determined Solar Cell Current Transport Efficiency Map

Amaury Delamarre, Laurent Lombez, Kentaroh Watanabe, Masakazu Sugiyama, Yoshiaki Nakano
and Jean-François Guillemoles

*Abstract*—A recently suggested reciprocity relation states that the current transport efficiency from the junction to the cell terminal can be determined by differentiating luminescence images with respect to the terminal voltage. The validity of this relation is shown experimentally in this paper, by comparison with simultaneously measured electrical currents and simulations. Moreover, we verify that the method is applicable under various light concentrations and applied voltages, which allows us to investigate the cell in relevant conditions. Results evidence several kind of series resistances affecting the current transport efficiencies. We show that the relative contribution of those different resistances to the loss in current collection is a function of the illumination intensity.

*Index Terms*— Characterization of PV, Luminescence, Photovoltaic cells, Transport efficiency, Electroluminescence

## I. Introduction

LUMINESCENCE characterization methods allow investigating numerous valuable properties of solar cells, such as the quasi-Fermi level splitting [1]–[3], External Quantum Efficiency (EQE) [4], [5], temperature [6], series resistances [7]–[11] and others. Since we can acquire images, mapping of cell properties can be recorded [7]–[13], which brings a much finer understanding about the cell mechanisms than global values. A recently introduced reciprocity relation [14] allows for the determination of current transport efficiencies maps, which we demonstrate experimentally in this communication.

The current transport efficiency $f_t$ at a position $(x, y)$ and its reciprocity relation are given by [14]:

$$f_t(x, y) = \left.\frac{\delta I_T}{\delta I_L(x, y)}\right|_{\delta V_T = 0} = \left.\frac{\delta V(x, y)}{\delta V_T}\right|_{\delta I_L = 0} \quad (1)$$

$I_T$ and $V_T$ are the terminal current and voltage, $I_L(x, y)$ the local light induced current collected at the junction, and $V(x, y)$ the local diode voltage. $f_t$ reflects collection loss on the carrier path from the junction to the external circuit, which results from various series resistances existing in a solar cell. Such a quantity is relevant for characterizing the performances of cells and modules. It can be measured with a Light Beam Induced Current (LBIC) setup which excites the cell locally and scans its surface [15]. However such a method can be time consuming, especially for high spatial resolution. Thanks to the reciprocity relation (1), it is possible to determine $f_t$ from luminescence images, which acquisition is rather straightforward.

Early applications of the reciprocity relation (1) were reported in [16]–[18]. However, the obtained transport efficiency maps were not compared with electrical currents, so that we cannot conclude on the validity of the method. We remove this ambiguity in this communication, by comparing the optical measurement and the cell terminal current, which provides a rigorous proof of concept. Moreover, one of the advantages of the method is that it should be applicable at different working points (i.e. illumination intensity and voltage), so that the cell can be investigated in relevant conditions close to real operation. This was experimentally investigated in [17] by applying different voltages at a given illumination. The results presented here take full advantage of the reciprocity relations by investigating complete voltage ranges under various illumination. We also note that the method was previously used on silicon [16]–[18] and CIGS cells [18], whereas we show its application to a III-V cell (GaAs).

## II. Theoretical elements

In order to demonstrate the relation between the luminescence emission and $f_t$, let us start by considering the emission $\phi(E, r)$, at an energy $E$ and position $r$, in the emitter,

A. Delamarre acknowledges support from the Japan Society for the Promotion of Science (JSPS).

A. Delamarre, L. Lombez, K. Watanabe, M. Sugiyama, Y. Nakano and J.-F. Guillemoles are with NextPV, LIA CNRS-RCAST/U. Tokyo-U. Bordeaux, 4-6-1 Komaba, Meguro-ku, Tokyo 153-8904, Japan (e-mail: delamarre@hotaka.t.u-tokyo.ac.jp).

A. Delamarre, K. Watanabe, and J.-F. Guillemoles are with the Research Center for Advanced Science and Technology, The University of Tokyo, 4-6-1 Komaba, Meguro-ku, Tokyo 153-8904, Japan.

L. Lombez is with the IRDEP, Institute of R&D on Photovoltaic Energy, UMR 7174, CNRS-EDF-Chimie ParisTech, 6 Quai Watier-BP 49, 78401 Chatou Cedex, France.

M. Sugiyama and Y. Nakano are with the Department of Electrical Engineering and Information Systems, The University of Tokyo, 7-3-1 Hongo, Bunkyo-ku, Tokyo 113-8656, Japan as described by the generalized Planck's law [19]:

$$\phi(E, r) = \alpha(E, r) \frac{n(E, r)^2}{4\pi^3 \hbar^3 c_0^2} \frac{E^2}{exp\left(\frac{E - \Delta\mu(r)}{kT}\right) - 1} \quad (2)$$

$\alpha$ is the absorption coefficient, $n$ the optical index, $\Delta\mu$ the quasi-Fermi level splitting, other constants having their usual significations. By integration along all optical paths leading to an emission at the surface position $(x, y)$ and an angle $\theta$, we obtain the luminescence flux $\Phi(E, x, y, \theta)$ [4]:

$$\Phi(E, x, y, \theta) \quad (3)$$

$$= \int T(E, r, x, y, \theta) \frac{\cos(\theta)}{n(E, r)^2} \phi(E, r). dr$$

$T(E, r, x, y, \theta)$ is the transfer probability, from the volume point $r$ to the surface point $(x, y)$, of a photon of energy $E$, resulting in an emission at $(x, y)$ with an angle $\theta$. We also introduce the black body radiation flux $\Phi_{bb}$:

$$\Phi_{bb}(E) = \frac{1}{4\pi^3 \hbar^3 c_0^2} \frac{E^2}{exp\left(\frac{E}{kT}\right) - 1} \quad (4)$$

And a function k(r, x, y) such that:

$$exp\left(\frac{\Delta\mu(r)}{kT}\right) = k(r, x, y) \ exp\left(\frac{qV(x, y)}{kT}\right) \quad (5)$$

Considering the cell without illumination, and in case the carrier lifetimes, the transport properties and the space charge region width do not depend on the working point, we note that k(r, x, y) can be identified with the carrier collection probability at the junction $f_c(r, x, y)$ as defined in [4]. Using the Boltzmann approximation, valid here as both E and E − qV ≫ kT, we rewrite the surface emission:

$$\Phi(E, x, y, \theta) \quad (6)$$

$$= \int T(E, r, x, y, \theta)\alpha(E, r) \cos(\theta) \, k(r, x, y). dr$$

$$\Phi_{bb}(E)exp\left(\frac{qV(x, y)}{kT}\right)$$

$$= K(E, x, y, \theta)\Phi_{bb}(E)exp\left(\frac{qV(x, y)}{kT}\right)$$

In case k(r, x, y) equals $f_c(r, x, y)$, $K(E, x, y, \theta)$ is the partial external quantum efficiency as defined in [4]. Differentiating the logarithm of the emission:

$$\frac{dln\big(\Phi(E, x, y, \theta)\big)}{dV(x, y)} = \frac{q}{kT} + \frac{dln\big(K(E, x, y, \theta)\big)}{dV(x, y)} \quad (7)$$

Equation (7) illustrates that the luminescence is modified in two manners under a variation of the junction voltage. Firstly, the emission can be varied uniformly in the volume of the cell, as is expressed by the first term on the right hand side. Secondly, the luminescence emission profile in the cell volume can be modified, as expressed by the second term, where the junction voltage dependence is borne by the k(r, x, y) function.

In the following, it will be interesting to determine in which cases the second term of equation (7) is negligible. We note that it equals zero when the quasi-Fermi level splitting gradient is independent on the working point, which should be the case of homojunctions made of high quality materials, such as the GaAs cell investigated in this paper. Nevertheless, we should not restrict the applicability of the method to such particular devices. Let us consider a solar cell at a working point so that

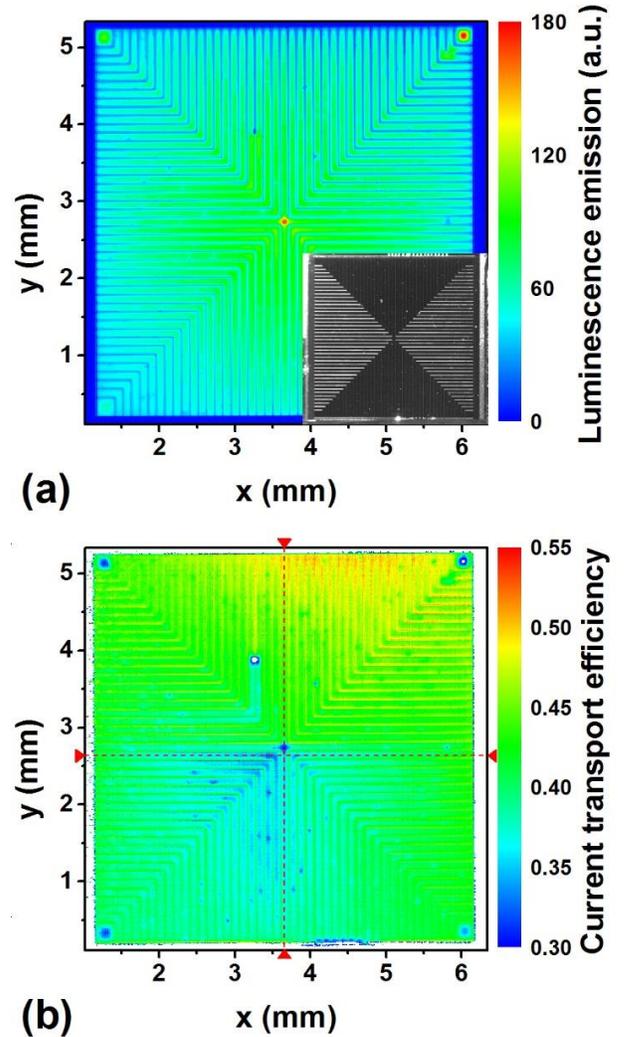

**(a)**

**(b)**

Fig. 1 (a) Luminescence image under 0.90 V applied voltage and 28 suns equivalent illumination. In inset is shown an image of the cell, on which we can see the connections to the external circuit on the upper edge of the cell. (b) Current transport efficiency map obtained with the images at 0.90 and 0.91 V. Along the dashed lines are taken the profiles displayed in Fig. 3.

$k(r, x, y) = f_c(r, x, y)$    (i.e.    $K(E, x, y, 0) = EQE(E, x, y)$). When the inequality $dEQE/dV \ll EQE \, q/kT$ holds (i.e. $\Delta EQE \ll EQE$ for a voltage variation of 26 mV), the second term of equation (7) is negligible. In such cases, we obtain by combining (1) and (7):

$$\frac{dln\big(\Phi(E, x, y, \theta)\big)}{d(qV_T/kT)} = f_t(x, y) \qquad (8)$$

Equation (8) shows that we can determine the carrier transport efficiency by monitoring the luminescence variation upon a small variation of the terminal voltage. We note that the derivation presented here differs from previous papers [14], [17], as we explicitly give the necessary condition $q/kT \gg dln(K(E, x, y, \theta))/(dV(x, y))$ in obtaining Equation (8). This derivation highlights that measuring $f_t$ by differentiating luminescence images may lead to wrong conclusions in case the local diode collection function $f_c$ [4] is not constant.

## III. EXPERIMENTAL SETUP

The experimental setup uses a Thorlabs CMOS camera mounted on a microscope for image recording, and a 2400 Keithley sourcemeter for current and voltage measurement and application. The illumination is obtained with a 532 nm laser from Coherent, expended for a homogeneous excitation. The laser light is filtered by a 780 nm long pass filter before images record. The solar cell is a pin GaAs solar cell grown by MOVPE, which description can be found in ref [20]. The illumination is quantified in number of suns, by comparison with the short-circuit current under AM 1.5 illumination. Illumination from 1 sun up to 47 suns equivalent are used, and voltages up to 1.25 V. The lowest voltage is limited by the setup sensitivity and depends on the illumination intensity. Images are recorded each 10 mV steps, with exposure time from 15 ms to 1 s, for the data in Fig. 2. Image averaging is possible to increase the signal to noise ratio, which was done with 50 images to obtain both maps in Fig. 1. Since the laser reflection is not completely filtered, a background subtraction is necessary. Background images are recorded under 0.5 V reverse bias.

## IV. RESULTS AND DISCUSSION

In Fig. 1(a) is displayed the cell luminescence under 28 suns equivalent illumination and 0.9 V applied voltage. With the image at 0.91 V and Eq.(8), the transport efficiency map in Fig. 1(b) is determined. After averaging over the cell area, the current transport efficiency is displayed as a function of the voltage and the illumination in Fig. 2(a). The same measurement was also performed at 0 sun (i.e. in electroluminescence, as in ref [16]), which resulted in values close to those obtained under 1 sun (not shown here).

Electrically, the global current transport efficiency of the cell can also be measured. This is done by subtracting the current under a given illumination and the current under the same illumination with a small increase (of about 3%). The result of this operation is normalized by its value at short-circuit, where it is expected that the carriers are collected without loss. For comparison, this electrical measurement of the transport efficiency is also shown in Fig. 2(a).

$f_t$ can be reduced by series resistances, since they result in an increased forward bias on the cell, implying additional recombination currents. This behavior was calculated in a simple model of 2 diodes $D_1$ (with ideality factor $n_1=1$ and saturation current $j_{01}=1.49e\text{-}15$ A/m²) and $D_2$ ($n_2=1.94$, $j_{02}=3.15e\text{-}7$ A/m²) in parallel with a current source $J_{gen}$, connected to the external circuit via a series resistance $R_S$ (Fig. 2(b)). Spread resistances were not included. The diodes ideality factors and dark currents were determined by fitting the dark IV characteristic of the cell. $f_t$ is calculated as a function of the applied voltage for each illumination, and the best correspondence with experimental data is found for a series resistance of 263 m$\Omega$.cm² (see Fig. 2).

As can be observed, we obtain a good agreement between the optical method, the electrical method and the simulation. This is not only a confirmation of the validity of Equation (1) by comparison with the cell current, but it also evidences that the reciprocity relations can be used at any operating point. Therefore the method allows investigating the cell under

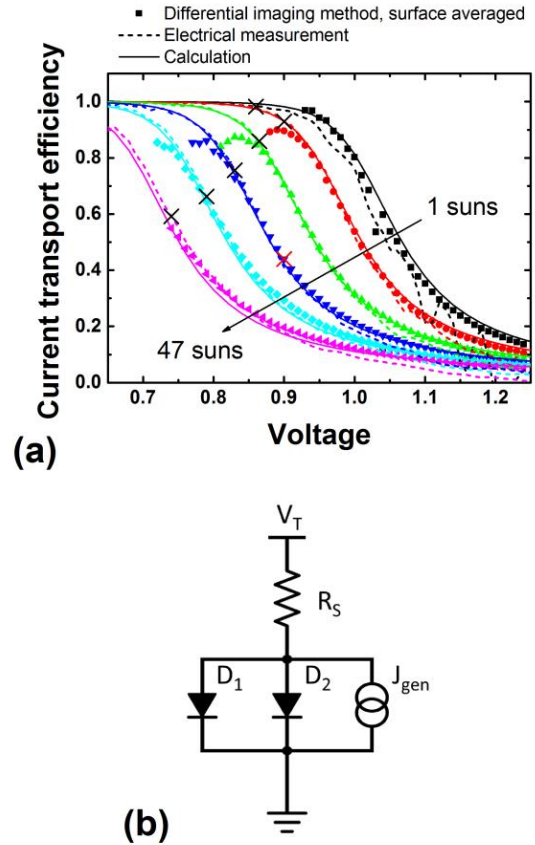

(a)

(b)

Fig. 2 (a) Current transport efficiencies under different applied voltages and illuminations (equivalent to 1, 9, 19, 28, 37 and 47 suns), obtained by three methods: (symbols) surface average of the transport efficiencies determined by differential imaging (application of Eq.(8)), (dashed lines) electrical measurement, (solid lines) calculations. The black crosses represent the current transport efficiency, determined by electrical measurements, at the maximum power points for each illumination. The red cross represents the acquisition point of data in Fig.(2). (b) Equivalent circuit used for the calculation of the collection efficiency.

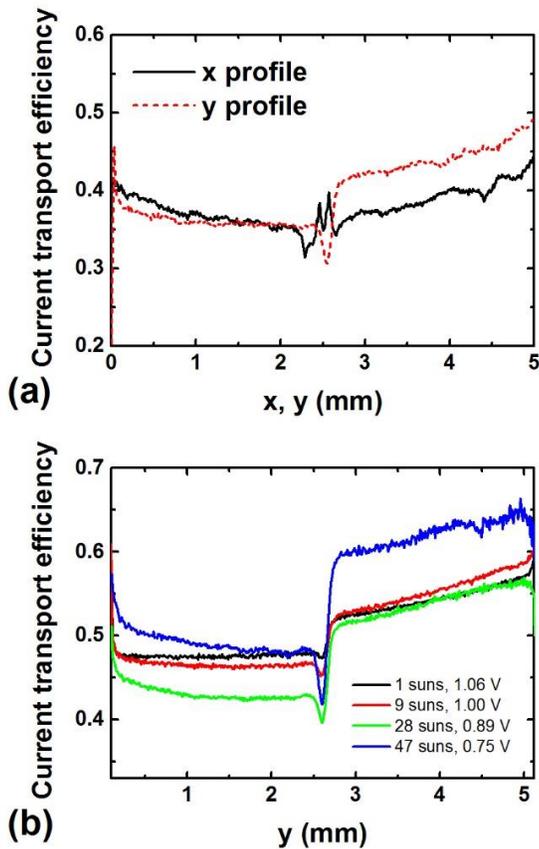

**(a)**

**(b)**

Fig. 3 (a) Transport efficiency profiles along the lines displayed in Fig. 1(b). (b) Transport efficiency profiles along the line parallel to the y-axis, for different illumination and voltages, resulting in a global transport efficiency around 0.5.

relevant condition close to real operation. For example, the contact pattern used here with small distance between the contact fingers is intended to be used under concentration, and we can see that the performance decreases for illuminations as low as 9 suns. This limits the maximum power point, whereas the open-circuit voltage continues to increase.

The origin of the transport efficiency decrease can be understood by taking advantage of the mapping capability of the method. On Fig. 1(b) we can first observe a clear difference between the upper and lower parts of cell, which is also shown by the efficiency profiles along the y-axis in Fig. 3(a) and (b). The whole contact pattern on the cell is connected to the external circuit only through bonding at the top edge of the cell (see inset picture in Fig. 1(a)), which explains this difference. The resistance of the contact finger itself decreases the collection, which is reflected by the v-shape profile along the x-axis in Fig. 3(a). A broken finger in the upper middle region is also seen to affect the cell performance. This defect occurs during lithography process. Eventually the window layer sheet resistance decreases the collection for carriers generated in between two contacts. This is better seen at the cell corners, where the distance to metallisation takes the highest values. We can also see that the maximum transport efficiency is not equal to 1 at any point on the surface, which means that another effect, not spatially distributed, is detrimental to the carrier collection. This can be the contact resistance between the GaAs window layer and the Ag/Au electrode. On practical cases, differentiating the loss mechanisms permits to identify the main one.

Here the maximum spatial variation is about 15%, whereas the global transport efficiency decreases from 100% to 10%, meaning that the largest loss is induced by the contact resistance. Indeed, the simulation carried out without spread resistance is sufficient to describe the cell behaviour. Considering the total metallisation surface, the series resistance used for simulation was 24 mΩ.cm², consistent with the value of 33 mΩ.cm² determined by TLM measurement with similar layers.

Fig. 3 (b) shows the collection profiles along the line parallel to the y-axis, for different illumination intensities and voltages, giving an average current transport efficiencies of about 0.5. As exposed above, we observe a sharp step at the middle of the cell, due to the specific pattern of the electrical contact. Although the average transport efficiencies are similar, the step increases with the illumination intensities from 0.04 to 0.12. This shows that the balance between the different mechanisms responsible for transport efficiency reduction depends on the working point investigated, and illustrates the relevance of the presented method and study.

## V. CONCLUSION

We demonstrated in this paper mapping of the current transport efficiency, using luminescence images with a method proposed in [14]. One of the advantages of this experiment is its validity at different working point, which was investigated and successfully compared with electrically measured current. Current transport efficiencies maps were measured on a GaAs cell, and different loss mechanisms were identified by spatial variation discussion. We could deduce that the major loss is induced by a non-spread resistance, which is possibly the metal – semiconductor contact resistance. The ability of measuring the cell properties at various working points is especially relevant in this case, since the contact pattern is intended for use under light concentration, where series resistance losses prevail and are strongly dependent on illumination.

## ACKNOWLEDGMENT

The authors thank Daiji Yamashita for the TLM measurements.